\documentclass[a4paper,11pt]{article}

\usepackage[english]{babel}

\usepackage[a4paper,top=2cm,bottom=2cm,left=3cm,right=3cm,marginparwidth=1.75cm]{geometry}

\usepackage{multirow}
\usepackage{amsmath}
\usepackage{graphicx}
\usepackage[colorlinks=true, allcolors=blue]{hyperref}
\usepackage{csquotes}
\usepackage[backend=biber,style=numeric,sorting=none]{biblatex}
\usepackage{authblk}
\usepackage{cuted}
\usepackage{caption}
\usepackage{algpseudocode}
\usepackage{algorithm}
\usepackage[most]{tcolorbox}
\usepackage{xcolor}
\usepackage{hyperref}

\newcommand{\figref}[1]{%
    \hyperref[{#1}]{\textbf{\textcolor{black}{\ref*{#1}}}}%
}

\providecommand{\keywords}[1]
{
  \small	
  \textbf{\textit{Keywords---}} #1
}

\title{In Situ Three-Dimensional Mapping 
of \\ Lithiation Heterogeneity 
in NMC Electrodes \\ using XANES Nanotomography
}

\author[1,2]{Tuan-Tu Nguyen}
\author[1,3]{Jiahui Xu}
\author[1,3]{Zeliang Su}
\author[5]{Vincent De Andrade}
\author[1,3]{Alejandro A. Franco}
\author[2]{Bruno Delobel}
\author[1,3]{Charles Delacourt}
\author[1,3,4,*]{Arnaud Demortière}

\affil[1]{Laboratoire de Réactivité et de Chimie des Solides (LRCS), CNRS UMR 7314, Université UPJV, Hub de l’Energie, Rue Baudelocque, 80039 Amiens Cedex, France.}
\affil[2]{Renault Technocentre, 78084 Guyancourt, France}
\affil[3]{Réseau sur le Stockage Electrochimique de l’Energie (RS2E), CNRS FR 3459, Hub de l’Energie, Rue Baudelocque, 80039 Amiens Cedex, France.}
\affil[4]{ALISTORE-European Research Institute, CNRS FR 3104, Hub de l’Energie, Rue Baudelocque, 80039 Amiens Cedex, France.}
\affil[5]{Advanced Photon Source, Argonne National Laboratory, Lemont, USA}

\affil[*]{corresponding author: arnaud.demortiere@cnrs.fr}

\begin{document}
\maketitle

\begin{tcolorbox}[colback=blue!6,colframe=gray!40,arc=4pt,boxrule=0.4pt, left=0.1pt, right=0.1pt, top=6pt, bottom=6pt]

\begin{abstract}
Spatial heterogeneity in lithiation complicates the interpretation
of electrode-averaged electrochemical measurements and the assessment
of local active-material utilization. Here, we use three-energy
X-ray absorption near-edge structure (XANES) nanotomography to
investigate lithiation distributions in NMC532 composite electrodes
with active-material loadings of 3 and 15~mg~cm$^{-2}$.
Tomographic acquisitions at 8.30, 8.35, and 8.43~keV combine
three-dimensional particle morphology with Ni K-edge chemical
sensitivity at a processed voxel size of 210~nm. The normalized
absorption contrast serves as a relative lithiation indicator,
without assigning absolute lithium concentrations. Measurements
were performed \emph{in situ} after 1~h of open-circuit relaxation
at selected electrochemical states. In the lower-loading electrode,
four successive states during interrupted discharge at
2~mA~cm$^{-2}$ reveal an increasing lithium-rich volume fraction
alongside core--shell-like distributions, nonuniform peripheral
enrichment, and particle-to-particle variability. In the
higher-loading electrode, two regions examined at 50\% depth
of discharge exhibit different lithiation-sensitive contrast
distributions despite sharing the same nominal electrochemical
state. These observations document heterogeneity within particles,
between particles, and between electrode regions, showing how
overall lithiation progression can coexist with spatially
nonuniform local states. Three-energy XANES nanotomography thus
provides a means of examining lithiation heterogeneity in dense
composite electrodes with limited spectral sampling. Correlative
phase-contrast imaging could further relate these distributions
to the arrangement of the carbon--binder domain and pore network.

\end{abstract}

\end{tcolorbox}

\keywords{NMC532 cathodes; three-energy XANES nanotomography; in situ chemical imaging; multiscale lithiation heterogeneity; three-dimensional electrode characterization; high-loading composite electrodes}

\newpage
\section{Introduction}

Lithium-ion batteries underpin the electrification of transport and the deployment of electrochemical energy storage, which keeps pressure on energy density, power capability, and durability \cite{ref1,ref2,ref3,ref4}. Reaching these targets depends not only on active material chemistry but also on the architecture of composite electrodes. Porous electrodes offer a large interfacial area for charge transfer while hosting pathways for ionic and electronic transport. Their performance therefore emerges from the coupled behavior of active particles, the electrolyte-filled pore network, and the carbon--binder domain (CBD). Densifying electrodes to raise their volumetric energy density makes uniform access to these pathways harder to maintain \cite{ref5}. Spatial variations in phase distribution and connectivity can then generate heterogeneous reaction rates, uneven active material utilization, and localized degradation \cite{ref6,ref8,ref9}.

The CBD embodies the competing transport requirements of composite electrode design. Conductive additives promote electronic connectivity, yet their distribution can also alter pore accessibility, ionic tortuosity, and the fraction of particle surface open to electrochemical exchange \cite{ref7b,ref7c}. Forouzan \emph{et al.} measured local electronic conductivity to relate electrode microstructure to transport properties \cite{ref7a}. Such approaches are informative. Translating structural heterogeneity into spatial variations in lithiation, however, requires complementary measurements of the local chemical state. Resolving these variations within particles and across electrode regions is decisive for separating particle-scale behavior from the collective response of the composite electrode.

Raman microscopy \cite{ref16}, X-ray diffraction (XRD) \cite{ref17,ref18}, X-ray absorption spectroscopy (XAS) \cite{ref19}, and spatially resolved X-ray absorption near-edge structure (XANES) measurements \cite{ref20,ref21,ref22} track structural transformations and redox evolution from complementary angles. Conventional ensemble measurements, however, average over heterogeneous populations, and two-dimensional transmission images integrate contributions along the beam direction. Without depth-resolved information, overlapping particles and internal chemical gradients are hard to tell apart. Three-dimensional imaging eases this ambiguity: it separates spatial contributions and ties local chemical states to the surrounding morphology \cite{ref23a,ref23b}.

X-ray tomography is now an established tool for characterizing electrode architecture, from particle morphology and pore connectivity to structural changes during electrochemical cycling \cite{ref10,ref11,ref12,ref13,ref14,ref15,ref15b}. Coupled with electrochemical measurements and complementary contrast mechanisms, it extends from structure toward the spatially resolved study of electrode reactions. Finegan \emph{et al.} demonstrated operando XRD computed tomography to resolve crystallographic heterogeneity within and between lithium manganese oxide particles \cite{ref24}. More recently, Vanpeene \emph{et al.} benchmarked X-ray nanotomography workflows across six battery materials and stressed the weight of acquisition conditions, image quality, and representative sampling in reliable morphological quantification \cite{Vanpeene2025NanoCT}. Spatial resolution must thus be weighed together with measurement fidelity and the representativeness of the analyzed volume.

Adding XANES to transmission X-ray microscopy (TXM) brings sensitivity to local oxidation state and chemical environment. Meirer \emph{et al.} established three-dimensional chemical imaging of the NiO-to-Ni conversion reaction, showing how spectroscopic tomography exposes chemical heterogeneity within complex electrode structures \cite{Meirer2011XANES}. Later studies carried spatially resolved XANES over to intercalation materials. Tan \emph{et al.} investigated nanoscale state-of-charge heterogeneity in polycrystalline nickel-rich cathodes, and Kimura \emph{et al.} used CT-XANES to visualize three-dimensional reaction distributions in a composite solid-state battery electrode \cite{ref25, ref26}. Uematsu \emph{et al.} recently combined two-dimensional TXM-XAFS (X-ray Absorption Fine-structure Spectroscopy) measurements, three-dimensional chemical imaging, and clustering analysis to examine heterogeneous charging and its relation to particle position within a positive electrode \cite{Uematsu2026ChemicalImaging}. Together, these studies establish chemical imaging as a bridge between local redox behavior and electrode organization.

Extending three-dimensional chemical imaging to dense composite electrodes is bound by a three-way trade-off between spatial resolution, chemical sensitivity, and acquisition time. Studies of individual particles, and of electrode configurations optimized for X-ray imaging, have yielded valuable insight into local reaction behavior; dense electrodes add constraints of their own, from X-ray attenuation to tomographic access \cite{ref20,ref27,ref28}. In XANES tomography, each additional energy lengthens the measurement and adds cumulative exposure, which can perturb the very electrochemical behavior under study \cite{Jousseaume2023BeamDamage,Rahman2025BeamDamage}. These constraints call for reduced-energy acquisition strategies that keep sensitivity to lithiation heterogeneity while limiting the number of tomographic scans.

Here, we pair TXM nanotomography with a three-energy acquisition scheme at the Ni K-edge to investigate lithiation heterogeneity in LiNi$_{0.5}$Mn$_{0.3}$Co$_{0.2}$O$_2$ composite electrodes. Measurements below, near, and above the edge yield a normalized absorption contrast sensitive to the local lithiation state. We examine electrodes with different active material loadings, including a dense electrode at 15~mg~cm$^{-2}$, in electrochemical cell configurations adapted to tomographic access. Three-dimensional datasets are collected \emph{in situ} at selected electrochemical states after current interruption and relaxation. They resolve lithiation distributions within secondary particles and across electrode regions. The approach tackles a practical problem, obtaining spatially resolved chemical contrast in dense electrodes from limited spectral sampling, and provides a basis for investigating persistent lithiation heterogeneity in relation to electrode microstructure.

\section{Experimental Section}

\subsection{Electrode composition and loading}

The two positive electrodes investigated share one composition: a mixture of
LiNi\textsubscript{0.5}Mn\textsubscript{0.3}Co\textsubscript{0.2}O\textsubscript{2}
(NMC) as AM, conductive carbon black and a mixture of polyvinylidene
fluoride (PVdF) (96\%w AM, 2\%w carbon black and 2\%w PVdF). They differ
in AM loading. MX-A, made in our laboratory, has a very low loading
(\emph{c.a.} 3 mg/cm²) and 30\% porosity after calendaring. MX-B is a
commercial electrode with a loading of 15 mg/cm² and 28\% porosity after
calendaring.

\subsection{Sample preparation for nanotomography}

Our setup requires X-rays to cross at least the full thickness of the
electrode, and the path length grows as the rotation angle sweeps through
{[}-150°,150°{]} (assuming that the glassy carbon windows, the Celgard
2500 and the Li foil are mostly transparent). As a transmission
technique, it needs enough X-rays leaving the sample to give sufficient
signal on the detector. Of the two electrodes used in this work, MX-B is
the problem: it is considerably thicker. Without preparation, its
rotation angle is limited to {[}-75°, 75°{]}. Sample preparation is
therefore decisive for MX-B, both for result quality and for limiting the
dead angle during acquisition.

Preparation starts from a ``free-standing'' electrode, obtained by simply
peeling the porous electrode material off the current collector foil. A
ZEISS laser dissection then cuts a precise pattern from the bulk
electrode (see Figure 1d). The pattern, together with the void around it,
shortens the X-ray path over a wide rotation range {[}-120°, 120°{]},
reducing the dead angle from 75° to 30°. Smaller dead angles make the
thicker MX-B usable for the experiment. To spare the region of interest
from laser damage, the laser was operated in pulse mode rather than
permanent mode.

\subsection{Electrochemical cell configurations and assembly}

Acquiring these datasets demands sufficient X-ray transmission and angular access through the electrochemical cell. Figures 1b, c, and e present the imaging setup and cell configurations. The custom cell comprises a stainless-steel cover, a movable stainless-steel disc, and a Teflon base; its geometry limits angular obstruction during tomography while preserving electrochemical function. The design has been described in detail previously.\cite{ref29}

Cell geometry also caps the accessible electrode thickness, since it sets both the X-ray path length and the angular range available for tomography. Two cell configurations were therefore used for the \emph{in situ} measurements. The AMPIX cell, developed at Argonne National Laboratory for synchrotron scattering and spectroscopy experiments,\cite{ref33,ref34} served for the thinner MX-A electrode. It supports electrochemical measurements, but its thickness and narrow angular aperture make it ill-suited to TXM tomography, increasingly so as electrode thickness grows. Its use was consequently restricted to MX-A, to keep the combined transmission and angular-access constraints in check.

For the higher-loading MX-B electrode (15~mg~cm$^{-2}$), the custom cell described above offered a geometry better suited to tomographic acquisition. Its reduced thickness and wider angular aperture were designed to improve X-ray access while maintaining electrical contact and compression of the electrode stack. Assembly follows coin-cell principles, adapted to sample rotation and transmission imaging. This configuration enabled the thicker electrode to be studied under the electrochemical conditions described in the Experimental Section.

The \emph{in situ} cell was assembled in an argon-filled glovebox. The
electrode was placed at the center of the cell, just above the glassy
carbon window. For MX-B, keeping the laser-cut patterns near the center
of the windows makes them easier to locate during acquisition. Lithium
foil served as the counter electrode, 1.2 M LiPF\textsubscript{6}
(dissolved in a solvent of 50\% EC and 50\% DMC by weight) as the
electrolyte, and Celgard 2500 as the separator. The cell was designed to
minimize the dead angle during acquisition. Cells were verified to cycle
stably under constant current. The same design is compatible with other
\emph{in situ} experiments, such as X-ray diffraction in transmission.

\subsection{Cycling protocol and interrupted-state measurements}

Electrochemical cycling was performed with a BioLogic multipotentiostat (SP200) within a voltage window of 2.5-4.3~V versus Li/Li$^{+}$. Both electrodes were first charged galvanostatically at C/10 to 4.3~V, then held potentiostatically at 4.3~V for 60~min.

Discharge followed at an areal current density of 2~mA~cm$^{-2}$ for MX-A and at a nominal rate of 1C for MX-B, both to a lower cutoff voltage of 2.5~V. MX-B discharge also included a potentiostatic hold at 2.5~V. For comparison, a reference specific capacity of 180~mAh~g$^{-1}$ gives nominal areal capacities of approximately 0.54 and 2.70~mAh~cm$^{-2}$ for MX-A and MX-B, respectively. Under this assumption, 2~mA~cm$^{-2}$ corresponds to approximately 3.7C for MX-A.

At selected discharge states, the current was interrupted and the cells relaxed at open circuit for 1~h before tomographic acquisition. The three-energy datasets were then collected at zero current for both electrode configurations. The maps therefore capture lithiation after relaxation, not instantaneous distributions during galvanostatic operation.

\subsection{Three-energy XANES tomographic acquisition}

X-ray imaging was performed on the transmission X-ray
microscope at beamline 32-ID-C of the Advanced Photon Source
(APS). Tomographic reconstruction used the total-variation (TV)
algorithm implemented in \emph{TomoPy}. Reference Ni K-edge XANES
spectra were acquired for the pristine electrode and for an
electrode charged to 4.3~V versus Li/Li$^{+}$, over
8.30--8.43~keV in 150 energy steps. These spectra set the three
acquisition energies: 8.30~keV below the edge, 8.35~keV in the
near-edge region, and 8.43~keV above the edge (Figure 1). At each
selected electrochemical state, after the relaxation period
described above, three tomographic datasets were acquired in
sequence, one per energy, with the cell at open circuit. Each
dataset comprised 721 projections over an angular range of
$-120^{\circ}$ to $+120^{\circ}$, with a field of view of
approximately $50 \times 50~\mu\mathrm{m}^{2}$. Each three-dimensional tomographic dataset required approximately 3~min of acquisition

Restricting acquisition to three energies cuts the number of
tomographic scans relative to more densely sampled XANES
measurements, such as the 12-energy approach of Nowack
\emph{et al.}\cite{ref20} Under otherwise identical conditions,
acquisition time and cumulative X-ray exposure fall accordingly.
The price is spectral: each voxel carries lithiation-sensitive
absorption contrast, not a complete XANES spectrum.

\subsection{Volume processing and inter-energy registration}

The reconstructed volumes had a native voxel size of approximately 30~nm. To reduce noise and facilitate inter-energy registration, they were binned using $7 \times 7 \times 7$ voxel blocks, yielding a processed voxel size of 210~nm. This choice was informed by Kimura \emph{et al.} \cite{ref26}, who reported limited additional noise reduction beyond $6 \times 6 \times 6$ voxel binning. The resulting voxel size defines the spatial sampling interval rather than the effective
chemical imaging resolution, which also depends on reconstruction quality, registration accuracy, and partial-volume effects. This sampling allows spatial contrast variations to be examined within NMC secondary particles with a median diameter of $D_{50} = 4.7~\mu$m.

Voxel-wise calculation of the normalized Ni K-edge contrast requires spatial alignment of the volumes acquired at the three energies. Registration was performed using \emph{elastix}, with an initial rigid transformation followed by nonrigid B-spline registration. The aligned volumes were then combined
to calculate the relative lithiation indicator defined below. Residual misalignment and interpolation may affect local contrast, particularly near particle boundaries, pores, and cracks. Because the three volumes were acquired sequentially, the resulting maps represent the electrode over a finite acquisition interval; registration corrects spatial mismatch but does not account for possible chemical evolution during
this interval.

\subsection{Normalized contrast and particle-level classification}

The registered tomographic volumes at the three selected energies
were combined voxel-wise into a normalized absorption contrast, $R$:
\begin{equation}
R = \frac{V_{\mathrm{edge}}-V_{\mathrm{pre}}}
         {V_{\mathrm{pre}}-V_{\mathrm{post}}}.
\label{eq:contrast}
\end{equation}
Here, $V_{\mathrm{pre}}$, $V_{\mathrm{edge}}$, and
$V_{\mathrm{post}}$ denote the reconstructed voxel values at
8.30, 8.35, and 8.43~keV, respectively. Dividing by the
pre-edge/post-edge difference scales the near-edge contrast
to the absorption step across the Ni K-edge.

The sensitivity of the Ni K-edge to electrochemical state is what
allows $R$ to serve as a relative lithiation indicator.
Reference-based linear-combination analysis has previously mapped
local electrochemical states from sparsely sampled XANES
spectra.\cite{ref20} The present three-energy measurement, by
contrast, does not independently establish a linear relationship
between $R$ and lithium stoichiometry. $R$ was therefore read as
a relative indicator, with no conversion to absolute lithium
content. Such a conversion would demand calibration against
reference states of known stoichiometry, validated over the
investigated lithiation range.

Secondary particles were segmented and labeled with the
\emph{Separate Objects} module in Avizo. Active-material voxels
fell into three contrast classes: lithium-poor ($R\leq0.6$),
moderately lithiated ($0.6<R\leq0.9$), and lithium-rich ($R>0.9$).
These labels describe relative lithiation-sensitive contrast;
they are neither calibrated stoichiometric intervals nor distinct
chemical phases. For each particle, the volume fraction of each
class was calculated relative to its total segmented
active-material volume, and likewise over the total segmented
active-material volume of each analyzed electrode region.

\begin{figure}[htbp]
    \centering
    \includegraphics[width=\textwidth]{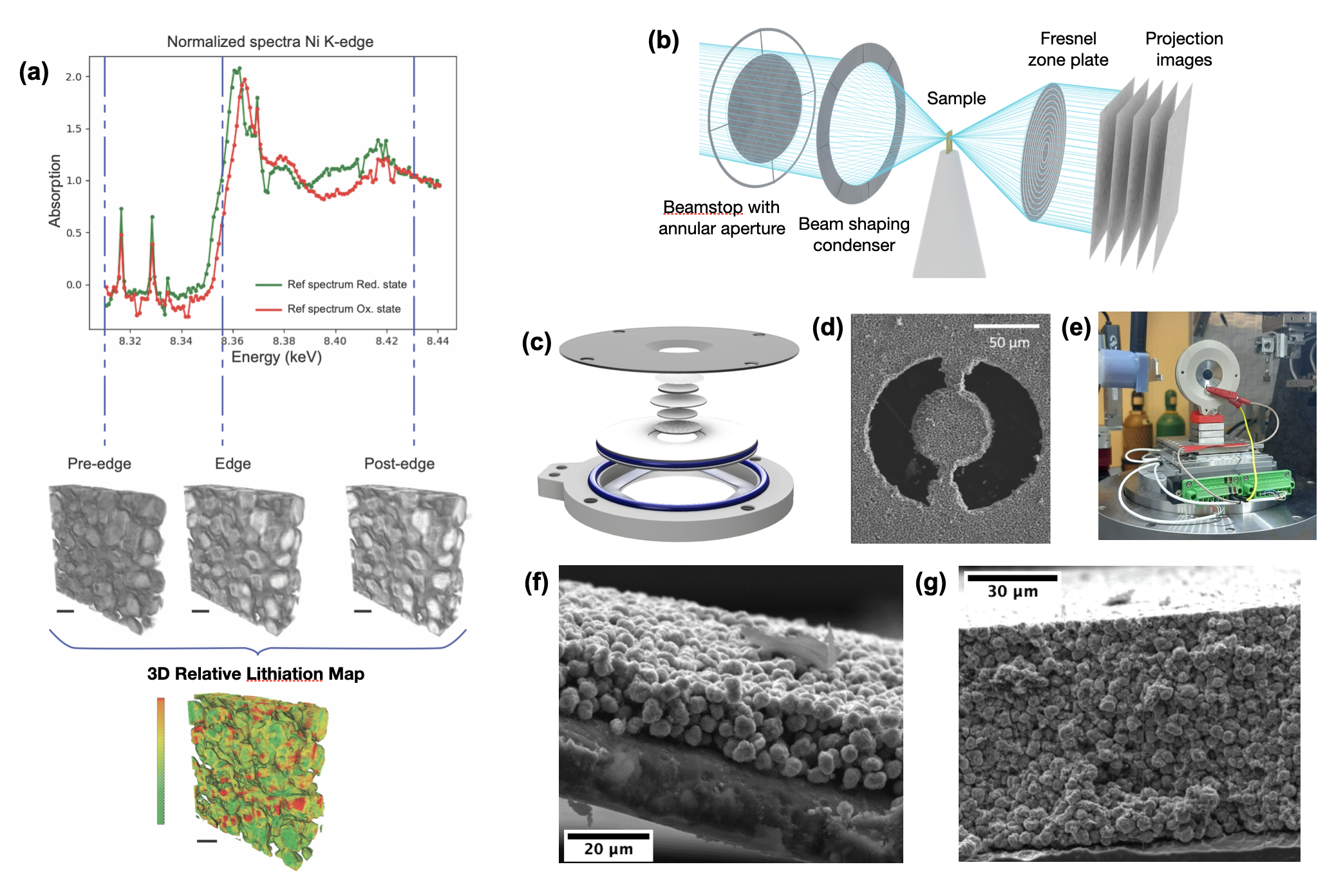}
   \caption{Experimental workflow for three-dimensional mapping of
lithiation-sensitive contrast in NMC532 electrodes.
(a) Reference Ni K-edge XANES spectra in reduced and oxidized
states set the three acquisition energies:
8.30~keV (pre-edge), 8.35~keV (near-edge), and 8.43~keV
(post-edge). Tomographic datasets acquired at these energies
are reconstructed, binned, and spatially registered, then
combined voxel-wise into the normalized contrast $R$,
yielding a three-dimensional map of relative lithiation
heterogeneity. Measurements are performed at open
circuit after a 1~h electrochemical relaxation period.
(b) Schematic of the transmission X-ray microscopy setup
used to collect projection images during sample rotation.
(c) Exploded view of the custom electrochemical cell.
(d) Laser-patterned electrode region designed to improve
X-ray transmission and angular access.
(e) Electrochemical cell mounted on the tomography stage.
(f, g) Cross-sectional scanning electron microscopy images
of MX-A and MX-B, with active-material loadings of
3 and 15~mg~cm$^{-2}$, respectively.}
\label{fig:1}
\label{strategy}
\end{figure}

\section{Results}

\subsection{Three-energy mapping of lithiation-sensitive contrast}

The three-energy approach couples Ni K-edge sensitivity with three-dimensional imaging of electrode morphology. Reference XANES spectra of the pristine and charged electrodes set 8.30, 8.35, and 8.43~keV as the pre-edge, near-edge, and post-edge acquisition energies, respectively (Figure 1). The near-edge signal tracks the local nickel environment as lithium is extracted and reinserted\cite{ref30}.

After reconstruction and inter-energy registration, the three volumes were combined into the normalized contrast $R$ defined in Equation~\ref{eq:contrast}. The resulting maps trace spatial variations in lithiation-sensitive contrast within and between secondary particles. Previous studies support Ni K-edge spectral features as trackers of electrochemical state;\cite{ref31,ref32} the fixed-energy contrast used here nonetheless remains a relative lithiation indicator, not a calibrated measurement of absolute lithium stoichiometry.

The approach was applied to two composite electrodes: MX-A, with an active-material loading of 3~mg~cm$^{-2}$, measured in an AMPIX cell, and MX-B, with a loading of 15~mg~cm$^{-2}$, measured in TuCell after laser patterning. Lithiation heterogeneity could thus be examined under two loading and imaging conditions. The electrodes also differed in cell assembly, preparation, and discharge protocol, so their comparison does not isolate the effect of loading alone.

Every map was acquired at open circuit after a 1~h relaxation period. What they reveal is the heterogeneity that survives current interruption.
The following sections follow its evolution within the repeatedly imaged MX-A region and its distribution across the sampled regions of MX-B.

\subsection{Electrode-scale lithiation heterogeneity in MX-A}

\begin{figure}[htbp]
    \centering
    \includegraphics[width=\linewidth,height=0.68\textheight,keepaspectratio]{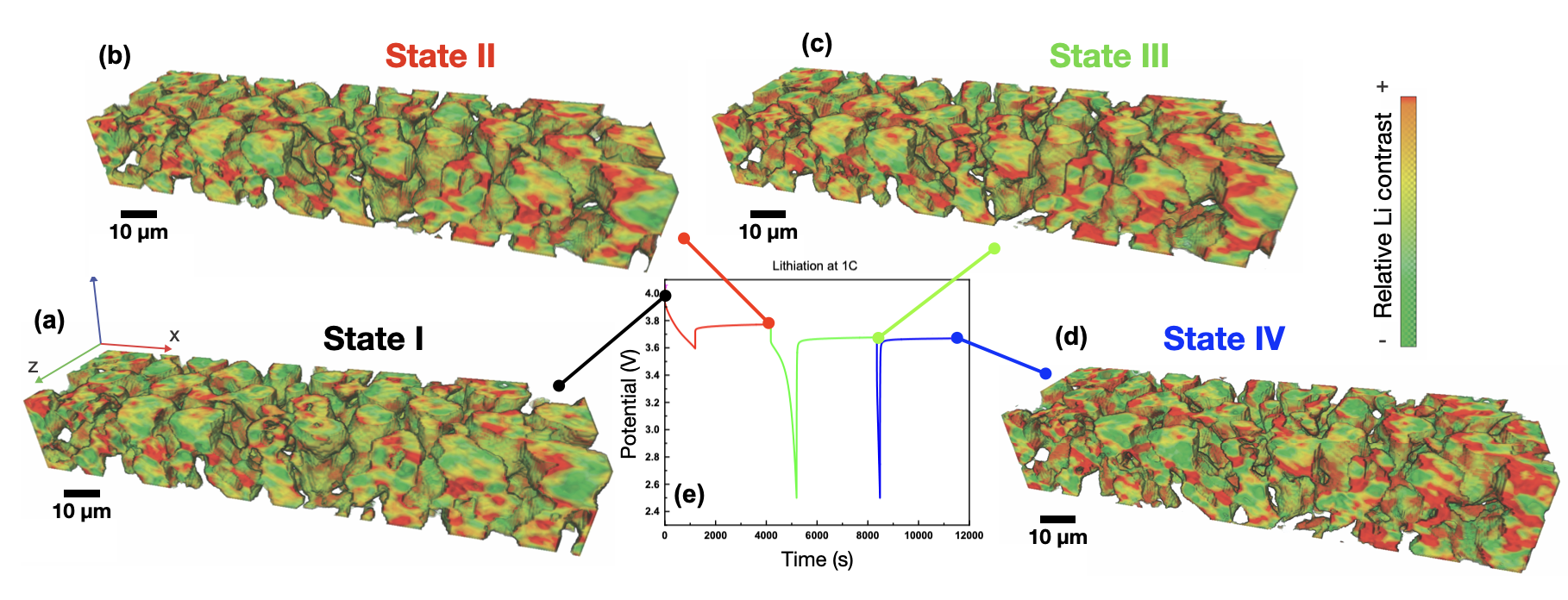}
        \caption{Three-dimensional evolution of lithiation-sensitive
    contrast in the MX-A electrode during interrupted discharge.
    (a-d) Maps of the same electrode region at successive
    states I-IV, acquired at open circuit after 1~h
    relaxation periods. The color scale represents relative
    lithiation contrast, with green and red indicating
    comparatively less- and more-lithiated regions,
    respectively. Red regions expand progressively and
    unevenly, within and between secondary particles.
    (e) Corresponding voltage profile during discharge steps
    at 2~mA~cm$^{-2}$ and subsequent relaxation periods.
    Colored markers and connecting lines identify the
    electrochemical states associated with the maps.
    Scale bars: 10~$\mu$m.}
    \label{fig:mxa_evolution}
    \label{fig:2}
\end{figure}

Repeated imaging of one MX-A electrode region allowed lithiation-sensitive contrast to be followed across successive discharge states.
Discharge steps ran at an areal current density of 2~mA~cm$^{-2}$, and each acquisition followed a 1~h relaxation period at open circuit.
Figure~\ref{fig:mxa_evolution} presents the voltage profile and the corresponding three-dimensional maps.

From state I to state IV, contrast rises overall, as regions displayed in red progressively expand. The rise is uneven. Higher-contrast regions cluster in the front-right portion of the displayed volume, while other regions retain comparatively lower contrast. The sampled electrode volume thus does not reach a spatially uniform lithiation state during the investigated discharge sequence.

The voxel-wise contrast distributions offer a complementary view. They shift toward higher values during discharge, in the direction expected for lithiation. Regions of different contrast nonetheless coexist: the overall increase is accompanied by persistent spatial heterogeneity. Directional profiles across the reconstructed volume (Figures~S4 and~S5) detail these variations.

Polarization becomes pronounced during the later discharge steps (Figure 2). Its possible link to cell assembly and electrical contact is discussed below. Because the maps were acquired after relaxation, they show heterogeneity retained after current interruption; they do not directly resolve local reaction rates during discharge.

\subsection{Intraparticle Li distributions and
particle-resolved analysis in MX-A}

\begin{figure}[htbp]
    \centering
    \includegraphics[width=\linewidth,height=0.68\textheight,keepaspectratio]{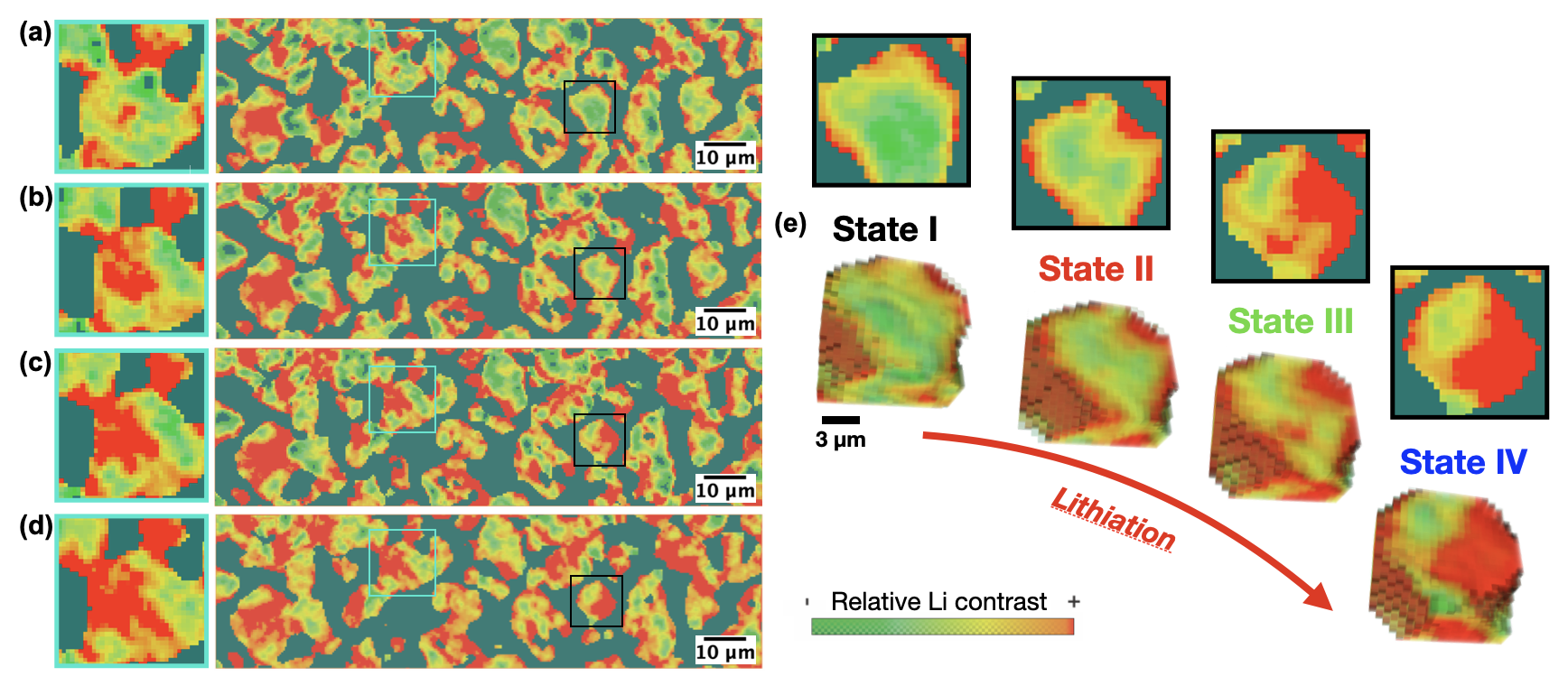}
    \caption{Intraparticle lithiation heterogeneity in the MX-A
    electrode during interrupted discharge.
    (a--d) Cross-sectional maps of the same electrode region
    at states I--IV, respectively, with enlarged views of
    the purple-boxed regions shown on the left.
    (e) Enlarged cross-sections of the black-boxed particle
    (top) and three-dimensional views of a selected particle
    (bottom), illustrating the evolution of intraparticle
    contrast across the four states. Green and red denote
    comparatively lower and higher lithiation-sensitive
    contrast, respectively. Lithium-rich peripheral regions
    surround less-lithiated interiors. All datasets were
    acquired at open circuit after 1~h relaxation
    periods. Scale bars in (a--d): 10~$\mu$m.}
    \label{fig:mxa_intraparticle}
    \label{fig:3}
\end{figure}

Beyond these regional variations, the
cross-sectional maps expose substantial contrast
heterogeneity inside individual secondary particles.
At successive discharge states, comparatively
lithium-rich rims enclose less-lithiated interiors,
giving core--shell-like distributions
(Figure~\ref{fig:3}a--d). Enlarged particle views show the higher-contrast
regions reaching further toward particle interiors
as discharge proceeds
(Figure~\ref{fig:3}e).
These patterns are consistent with surface-accessible
lithium insertion followed by redistribution within
the particles, as discussed in previous
studies.\cite{ref16,ref28,ref35}
Relaxed-state measurements, however, cannot
resolve the motion of a reaction front during
current flow.

Nor is the peripheral contrast uniform along
particle boundaries. Lithium-rich domains occupy
selected stretches of the periphery instead of
forming a continuous shell of even contrast.
A purely radial description therefore falls short.
The asymmetry may reflect differences in local
transport access or particle geometry; the maps
alone cannot identify the dominant mechanism,
nor classify lower-contrast surface regions as
electrochemically inactive.

\begin{figure}[H]
    \centering
    \includegraphics[width=\linewidth,height=0.68\textheight,keepaspectratio]{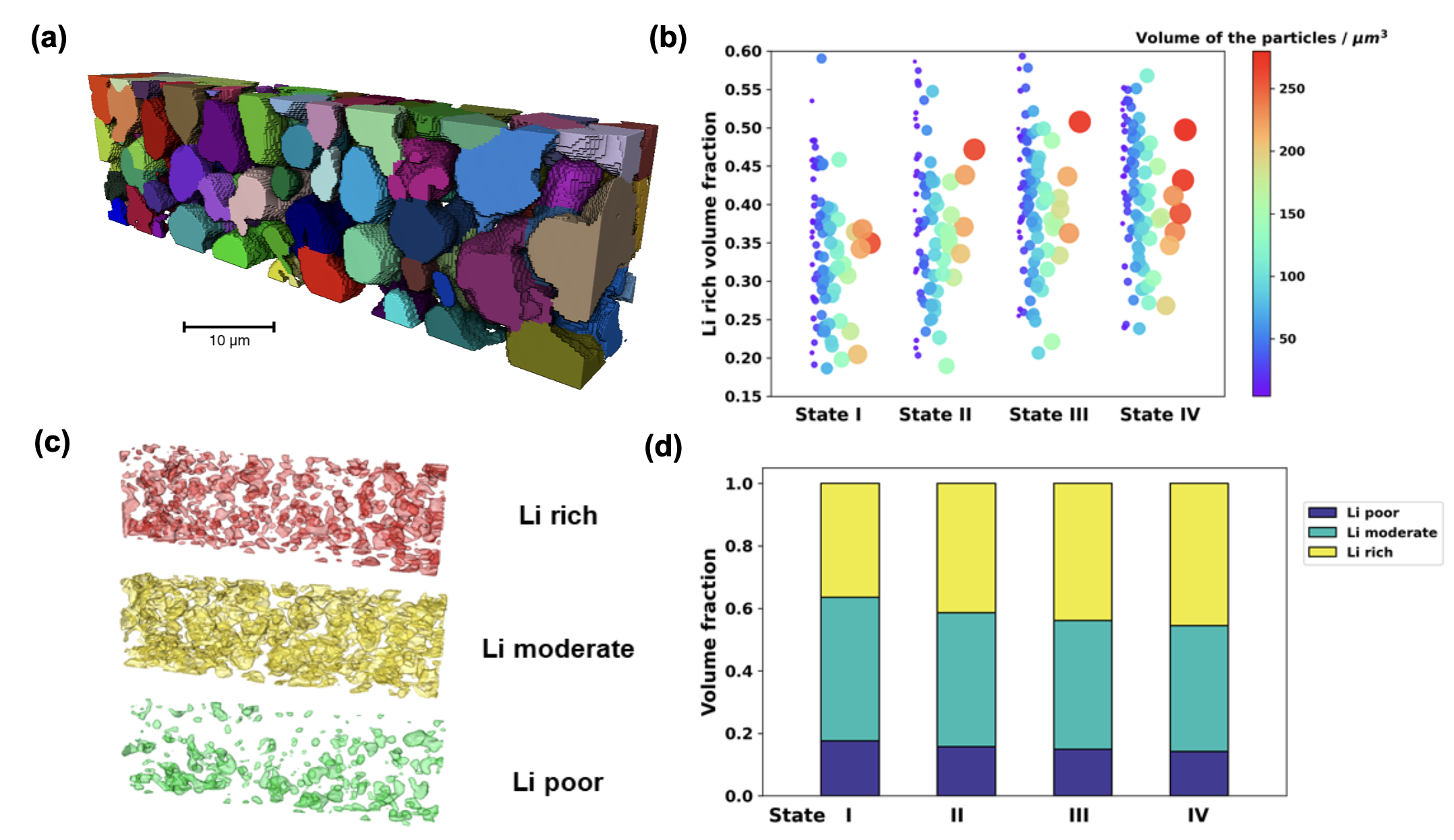}
    \caption{Particle-resolved quantification of lithiation
    heterogeneity in the MX-A electrode.
    (a) Three-dimensional rendering of segmented secondary particles, with distinct colors identifying individual particles. Scale bar: 10~$\mu$m.
    (b) Lithium-rich volume fraction within individual
    particles at discharge states I--IV. Each marker
    represents one particle; marker color and size indicate
    its segmented volume.
    (c) Separate three-dimensional renderings of the
    lithium-rich, moderately lithiated, and lithium-poor
    contrast classes.
    (d) Volume fractions of the three classes at states
    I--IV, normalized to the total segmented active-material
    volume in the analyzed region. The classes are defined from the normalized XANES contrast as lithium-poor, moderately lithiated, and lithium-rich.}
    \label{fig:mxa_statistics}
    \label{fig:4}
\end{figure}

To extend these observations beyond selected
particles, secondary particles were individually
segmented and labeled
(Figure~\ref{fig:4}a).
Their lithium-rich volume fractions increase
overall from state I to state IV, across the
sampled range of particle volumes
(Figure~\ref{fig:4}b).
The spread at each state reveals particle-to-particle
variability but no systematic dependence on
particle size.

The three contrast classes defined in the
Experimental Section give a complementary
picture of lithiation within the analyzed
volume. Figure~\ref{fig:4}c shows their spatial
distributions; Figure~\ref{fig:4}d summarizes
their volume fractions.
During discharge, the lithium-rich fraction
grows as the lithium-poor fraction shrinks.
These changes quantify the redistribution of
relative lithiation contrast without assigning
absolute lithium stoichiometries to the
individual classes.

Segmented particle volume also increases slightly.
Attributing this to particle expansion is
uncertain: close interparticle contacts and
internal cracks complicate segmentation.
More broadly, reconstruction regularization,
partial-volume effects, and inter-energy
registration can all shape peripheral contrast.
Their bearing on intraparticle gradients is
addressed in the Discussion.

\subsection{Evolution of lithiation-sensitive contrast
during cycling of MX-B}

\begin{figure}[htbp]
    \centering
    \includegraphics[width=\linewidth,height=0.68\textheight,keepaspectratio]{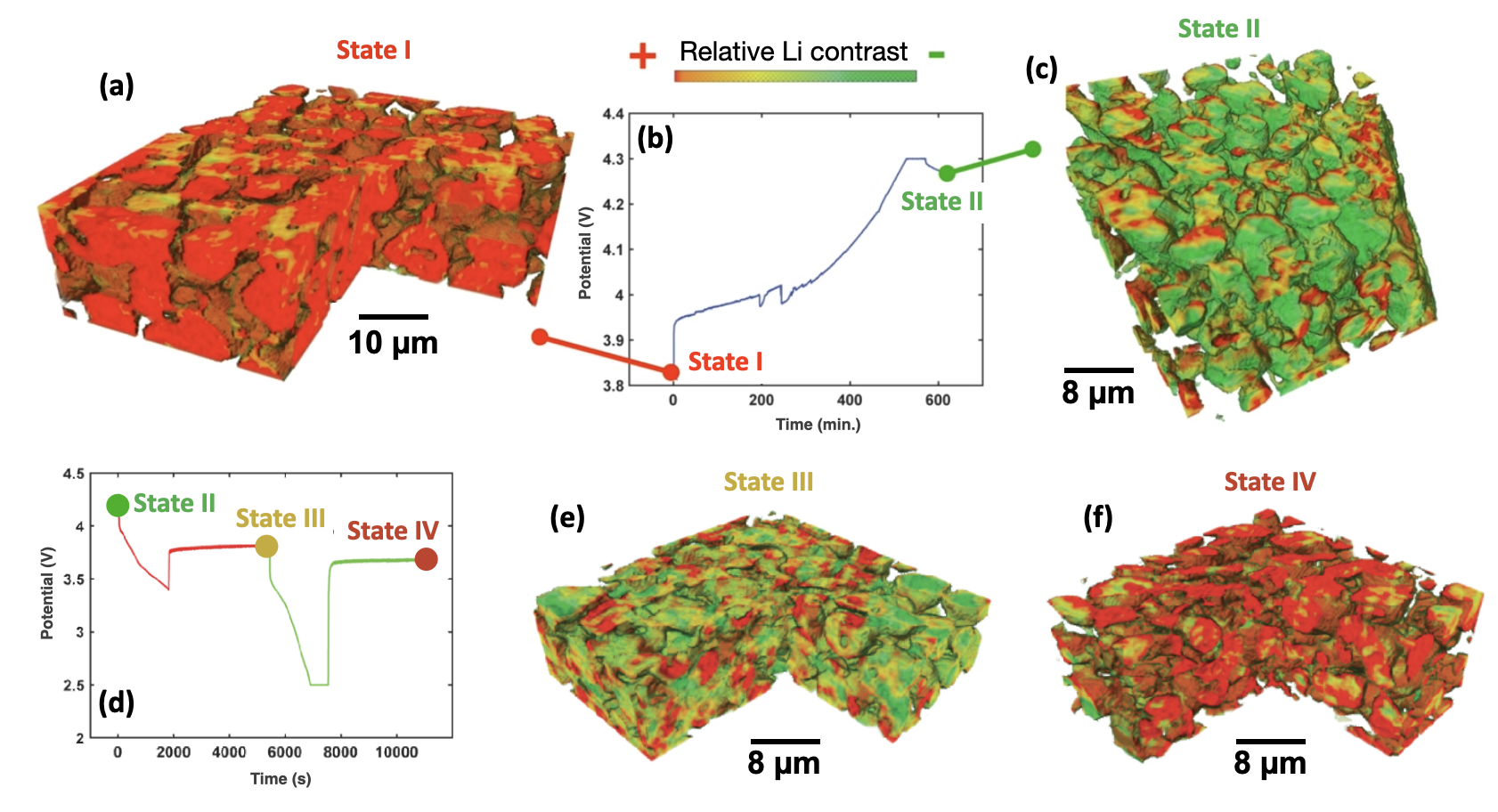}
    \caption{Evolution of lithiation-sensitive contrast during
    cycling of the high-loading MX-B electrode
    (15~mg~cm$^{-2}$).
    (a) Three-dimensional map of the pristine electrode
    (state I).
    (b) Voltage profile during galvanostatic charging at
    C/10 to 4.3~V versus Li/Li$^{+}$, followed by a
    60~min potentiostatic hold and open-circuit relaxation.
    (c) Map acquired after charging and relaxation
    (state II).
    (d) Voltage profile during discharge at a nominal
    rate of 1C, interrupted for relaxation and imaging
    at state III, followed by discharge to 2.5~V,
    a potentiostatic hold, and final relaxation.
    (e, f) Maps acquired at 50\% depth of discharge
    (state III) and after the final discharge and
    relaxation sequence (state IV), respectively.
    Colored markers identify the electrochemical states
    associated with the maps. Post-charge and discharge
    maps were acquired at open circuit after
    1~h relaxation periods. Red and green indicate
    comparatively higher and lower relative lithiation
    contrast, respectively. The sampled region differs
    between some states; the maps therefore do not
    track an identical particle population throughout
    cycling. Scale bars: 10~$\mu$m in (a) and
    8~$\mu$m in (c), (e), and (f).}
    \label{fig:mxb_evolution}
    \label{fig:5}
\end{figure}

The higher-loading MX-B electrode tested whether
three-energy tomography carries over to a dense
composite electrode with an active-material
loading of 15~mg~cm$^{-2}$.
The pristine-state map is predominantly
lithium-rich
(Figure~\ref{fig:mxb_evolution}a).
After charging at C/10 to 4.3~V, a 60~min
potentiostatic hold, and open-circuit relaxation,
the sampled volume shows lower and more spatially
variable contrast
(Figure~\ref{fig:mxb_evolution}b,c).
This change is consistent with lithium extraction.
The pristine and charged maps, however, come from
different electrode positions, so they compare
two sampled populations, not identical particles
over time.

During the subsequent discharge at a nominal rate
of 1C, maps were acquired after relaxation at 50\%
depth of discharge and after discharge to 2.5~V
followed by a potentiostatic hold
(Figure~\ref{fig:mxb_evolution}d--f).
Higher-contrast regions become more prevalent,
consistent with lithium reinsertion.
Particle movement and detachment prevented repeated
imaging of identical regions and left fewer
accessible measurement states than for MX-A
(Figure~S1).

The contrast histograms and class fractions in
Figure~S2 complement these spatial observations.
In its overall lithiation indicator, the final
discharged volume approaches the pristine one.
Its contrast distribution remains broader.
This difference is consistent with residual
lithiation heterogeneity after cycling and relaxation.
Its interpretation must nevertheless account for
the different sampled regions and possible
differences in reconstruction quality.
The measurements thus show sensitivity to
chemical-state variations in the higher-loading
electrode, with only limited temporal tracking
of individual particles.

\subsection{Regional Li heterogeneity in MX-B
at a common discharge state}

\begin{figure}[htbp]
    \centering
    \includegraphics[width=\linewidth,height=0.68\textheight,keepaspectratio]{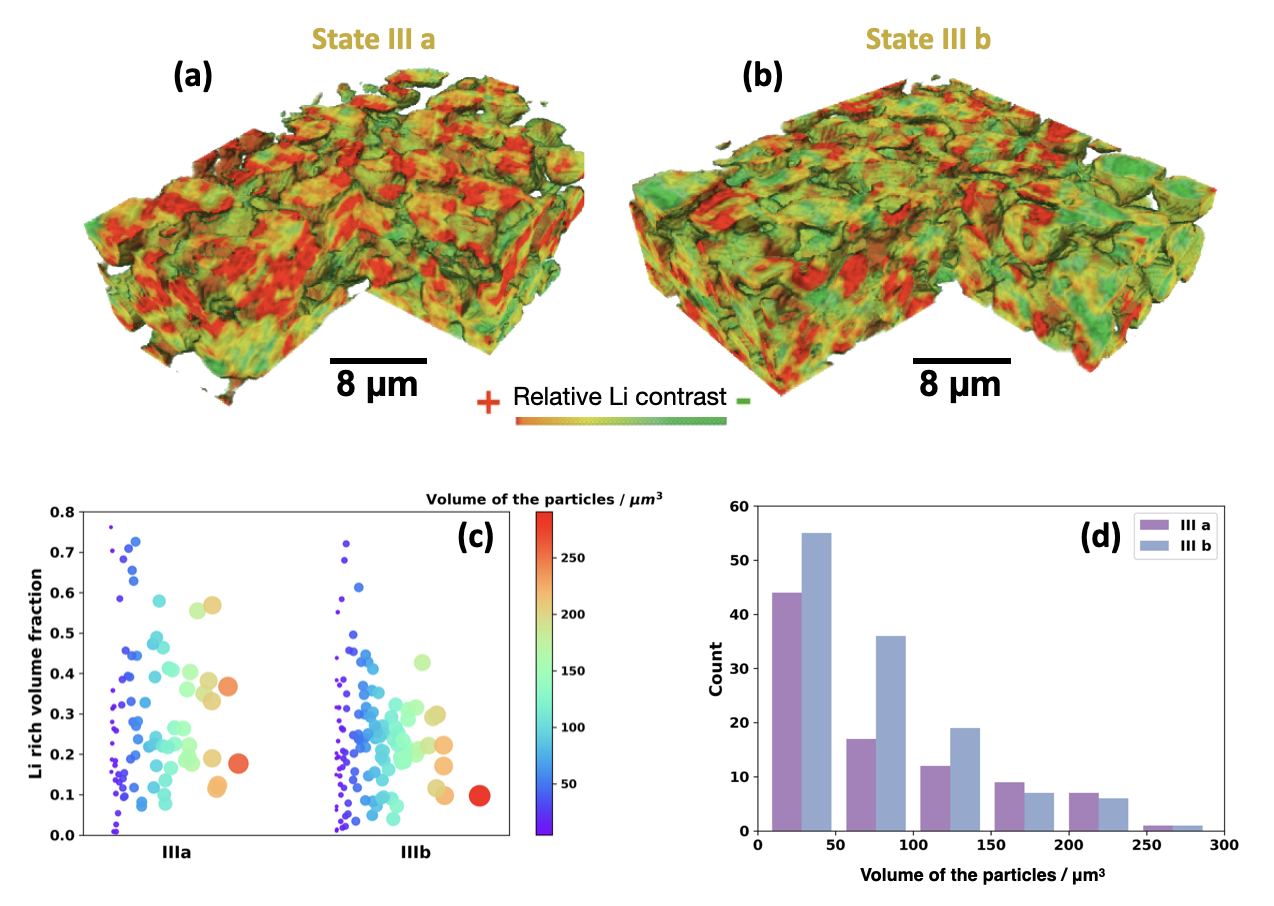}
    \caption{Regional lithiation heterogeneity in the MX-B
    electrode at 50\% depth of discharge.
    (a, b) Three-dimensional maps of two electrode regions,
    denoted IIIa and IIIb, examined at the same nominal
    discharge state after a 1~h open-circuit relaxation
    period. Red and green indicate comparatively higher
    and lower relative lithiation contrast, respectively.
    Scale bars: 8~$\mu$m.
    (c) Particle-resolved lithium-rich volume fractions
    in the two regions. Each marker represents one
    segmented secondary particle, with marker color
    and size indicating its volume. The lithium-rich
    fraction is defined as the fraction of particle
    volume satisfying $R>0.9$.
    (d) Histograms of segmented particle volumes in
    regions IIIa and IIIb, expressed as particle counts
    per volume bin. The comparison reveals regional
    differences in lithiation-sensitive contrast and
    particle populations without establishing a causal
    relationship between particle size and lithiation.}
    \label{fig:mxb_regions}
    \label{fig:6}
\end{figure}

Spatial variability within MX-B was probed further
by comparing two electrode regions at the same
nominal discharge state.
Regions IIIa and IIIb were imaged at 50\% depth
of discharge after open-circuit relaxation
(Figure~\ref{fig:mxb_regions}a,b).
Region IIIa holds more extensive lithium-rich
domains. Region IIIb shows more lower-contrast
regions.

Particle-resolved analysis compares the two
populations from another angle
(Figure~\ref{fig:mxb_regions}c).
Both regions show substantial particle-to-particle
variability, with overlapping distributions of
lithium-rich volume fraction.
Their differences show that a single nominal
electrode discharge state can span distinct
local lithiation responses.

The particle-volume histograms
(Figure~\ref{fig:mxb_regions}d) describe the
populations sampled in each region.
Because they report absolute counts, their
comparison mixes the number of segmented
particles with their volume distribution.
These data alone do not make particle size
the origin of the regional contrast differences.
Local connectivity, electrolyte access, and
sample preparation may also contribute, as
considered in the Discussion.

Slice-resolved contrast-class profiles along the
reconstruction $Z$-axis appear in Figure~S3.
They characterize spatial variations within the
sampled volumes; reading them as through-thickness
variations would require a known orientation of
the electrode relative to the reconstruction axes.
Measurements after relaxation also cannot
determine whether electrolyte transport limited
the reaction during the preceding discharge steps.

\section{Discussion}

\textbf{Lithiation heterogeneity across particle and electrode scales.}\\
The reconstructed volumes show lithiation-sensitive contrast variations within secondary particles, between particles, and between electrode regions. In MX-A, the increase in the lithium-rich volume fraction during discharge is accompanied by nonuniform spatial distributions, including differences between particle interiors and peripheral regions. In MX-B,
regions IIIa and IIIb exhibit different contrast distributions at the
same nominal discharge state. 
These observations illustrate how an
electrode-level electrochemical state can encompass a range of local
lithiation states. Spatially resolved XANES measurements have previously revealed
intraparticle state-of-charge heterogeneity in polycrystalline,
nickel-rich layered oxides\cite{ref25}. The present results extend this
discussion to the spatial distributions observed in the investigated
NMC532 composite electrodes. Their relevance lies in the coexistence of
heterogeneity at several scales: an increase in the overall lithiation
indicator does not imply that all particles, or all regions within a
particle, evolve uniformly. 

Here, these distributions are inferred from
relative Ni K-edge contrast rather than independently calibrated lithium
concentrations. The regional differences observed in MX-B are relevant to the broader
question of reaction uniformity in composite electrodes. Combined
imaging and modeling of thick NMC811 cathodes has linked heterogeneous reaction distributions to the organization and evolution of ionic and electronic transport networks \cite{Zhang2023ThickCathode}. Our maps
provide spatial information relevant to this problem, although they do
not identify the transport process responsible for the observed
differences. Moreover, MX-A and MX-B differ in cell configuration,
sample preparation, and discharge conditions as well as loading.
Their comparison therefore does not establish a loading-dependent
increase in heterogeneity.

\textbf{Core-shell-like distributions and nonuniform peripheral contrast.}\\
The core-shell-like patterns observed in MX-A are consistent with differences in lithiation between the peripheral and interior regions of secondary particles. However, the peripheral contrast is not uniform around every particle. The observed distributions therefore suggest deviations from an idealized, radially symmetric description, rather than establishing a uniformly advancing lithiation front. Several factors could contribute to these patterns. Mistry
\emph{et al.}\cite{ref36} showed through modeling that particle
asphericity can generate nonuniform intercalation. Jiang
\emph{et al.}\cite{ref35} demonstrated that particle
detachment from the carbon--binder domain can alter the distribution
of local electrical resistance. These studies provide possible
interpretive frameworks for spatially heterogeneous lithiation.
The present measurements, however, do not separate the effects of
particle geometry, electronic connectivity, electrolyte accessibility,
and intraparticle transport.

Higher contrast near some internal pores also raises the possibility
of lithium access through internal interfaces, consistent with
previous reports of grain-boundary penetration.\cite{ref37}
Such spatial associations do not establish that these pores are
electrolyte-connected or electrochemically active. Likewise, lower
contrast near a particle boundary cannot be interpreted directly as
an inactive surface. The maps record a local chemical state resulting
from the preceding electrochemical history and subsequent
redistribution, rather than the instantaneous interfacial reaction
rate.

\textbf{Interpretation of heterogeneity after current interruption.}\\
The measurement protocol is central to the interpretation of these
distributions. Post-charge and discharge acquisitions were performed
at zero applied current after a 1~h relaxation period. The maps
therefore characterize states observed after relaxation; they do not
directly capture the distributions present during galvanostatic
cycling. Further redistribution may also occur during the sequential
acquisition of the three tomographic datasets.

Recent experiments and simulations on faceted nickel-rich NMC
particles have highlighted the contribution of interparticle exchange
to lithium redistribution during rest.\cite{Sedajova2026Interparticle}
Although the particle morphology and composition differ from those
investigated here, this work emphasizes that zero external current
does not necessarily imply an absence of local lithium redistribution.
Accordingly, the heterogeneity observed after 1~h does not establish
either complete equilibration or irreversible particle deactivation.

The measured distributions should also be considered alongside the
electrochemical response and sampling conditions. In MX-A, the
pronounced polarization may include contributions from stack
compression and interfacial contact in the AMPIX configuration,
although these contributions were not isolated. In MX-B, changes in
the sampled region limit direct comparisons between successive
states. In particular, differences between pristine and discharged
contrast distributions cannot be attributed exclusively to cycling.
The comparison between regions IIIa and IIIb instead documents
spatial variability at a common nominal discharge state.

\textbf{Implications and scope of three-energy chemical imaging.}\\
These observations underline the value of describing lithiation
through spatial distributions alongside electrode-averaged
electrochemical quantities. Particle-averaged values can conceal
internal variations, while electrode-averaged values can conceal
differences between particles and regions. Three-dimensional imaging
provides complementary information by locating these variations
within the reconstructed morphology.

For modeling, the observed patterns motivate consideration of
departures from idealized spherical particles and uniform local
boundary conditions. They do not, by themselves, invalidate
homogenized electrode descriptions, whose suitability depends on the
scale and observable of interest. Nor do they determine an
electrochemically active surface fraction or identify a unique
transport limitation. Their contribution is to document spatial
features that such descriptions may need to account for when local
lithiation distributions are the target observable.

The three-energy acquisition provides this information with reduced
spectral sampling, but its contrast remains a relative lithiation
indicator. The lithium-poor, moderately lithiated, and lithium-rich
classes are operational categories, not distinct phases or absolute
stoichiometric intervals. The processed voxel size of around 200~nm is also
a sampling interval rather than an independently established chemical
resolution. Reconstruction, registration, partial-volume effects, and
segmentation uncertainty warrant particular caution near particle
boundaries, pores, and cracks.

\section{Conclusion}

Three-energy XANES nanotomography enabled three-dimensional
characterization of lithiation heterogeneity in NMC532 composite
electrodes containing 96~wt\% active material, with loadings of
3 and 15~mg~cm$^{-2}$ and porosities of 30\% and 28\%, respectively.
Acquisitions at 8.30, 8.35, and 8.43~keV combined particle morphology
with Ni K-edge chemical sensitivity using three tomograms per
measurement state. A processed voxel size of 210~nm allowed
lithiation-sensitive contrast variations to be examined within
secondary particles with a median diameter of 4.7~$\mu$m.

In the lower-loading MX-A electrode, four successive states during
interrupted discharge at 2~mA~cm$^{-2}$ showed an increase in the
lithium-rich volume fraction and a decrease in the lithium-poor
fraction. This overall progression was accompanied by spatially
nonuniform lithiation: comparatively lithium-rich peripheral
regions coexisted with less-lithiated particle interiors, while
peripheral enrichment varied around individual particles.
Particle-resolved analysis further revealed differences in
lithium-rich volume fraction between particles at each examined
state. Together, these results show that the evolution of an
electrode-averaged lithiation indicator can conceal both internal
particle gradients and variability within the particle population.

The higher-loading MX-B electrode extended the measurements to
15~mg~cm$^{-2}$. Its contrast distributions changed consistently
with lithium extraction during charge and reinsertion during
discharge. At 50\% depth of discharge, regions IIIa and IIIb
displayed different distributions, with more extensive
lithium-rich domains in region IIIa. This regional comparison
demonstrates that a common nominal electrochemical state does
not uniquely describe the local lithiation state throughout
the sampled electrode. The different cell configurations,
sample preparations, and discharge conditions nevertheless
prevent attributing differences between MX-A and MX-B to
electrode loading alone.

The contribution of this study is therefore to connect
lithiation-sensitive distributions across intraparticle,
interparticle, and regional scales within composite electrodes.
The maps were acquired after 1~h of open-circuit relaxation and
describe the states observed after this interval, rather than
instantaneous reaction distributions under current. Although
the normalized contrast is a relative indicator rather than an
independently calibrated absolute lithium concentration, it
provides spatial information complementary to the global
electrochemical response and reveals heterogeneity that is
not captured by a single mean lithiation value.

A promising extension is correlative XANES and X-ray
phase-contrast nanotomography of the same electrode volume.
With suitable spectral calibration, XANES could provide
estimates of local lithium content, while phase contrast could
improve visualization of the weakly absorbing carbon--binder
domain (CBD), particle--CBD contacts, and surrounding pores.
Registering these complementary datasets would enable direct
spatial comparisons between lithiation patterns and the local
electrode architecture. Such measurements could help determine
whether the observed peripheral and regional heterogeneities
are associated with differences in particle connectivity or
pore accessibility, and how these associations evolve during
cycling and relaxation.

\section*{Acknowledgements}

The authors acknowledge partial financial support from the French
National Association for Research and Technology (ANRT) and the
French National Research Agency (ANR) through the DynaMoBat project
(ANR-22-CE42-0025). T.-T.N. acknowledges Renault Group for supporting
his doctoral research. J.X. and A.A.F. acknowledge funding from the
European Research Council (ERC) under the European Union's Horizon
2020 research and innovation programme (grant agreement No.~772873,
``ARTISTIC'' project). This research used resources of the Advanced
Photon Source, a U.S. Department of Energy (DOE) Office of Science
User Facility operated for the DOE Office of Science by Argonne
National Laboratory under Contract No.~DE-AC02-06CH11357.

\section*{Declaration of Competing Interest}

The authors declare no conflict of interest.

\printbibliography
\end{document}